\documentclass[12pt]{article}
\usepackage{amscd,amssymb,amsmath,latexsym,enumerate}
\usepackage[mathscr]{euscript}
\usepackage{mathrsfs}
\usepackage{epsfig}
\usepackage{fancybox}
\usepackage{verbatim}
\usepackage{tikz}
\usepackage{tikz-cd}
\usetikzlibrary{arrows}

\usepackage{todonotes}
\usepackage{multicol}
\usepackage{graphicx}

\usepackage{color}

\usepackage{xcolor}
\definecolor{MyBlue}{cmyk}{1,0.13,0,0.63}
\definecolor{MyGreen}{cmyk}{0.91,0,0.88,0.52}
\newcommand{\mylinkcolor}{MyBlue}
\newcommand{\mycitecolor}{MyGreen}
\newcommand{\myurlcolor}{black}

\usepackage{hyperref}
\hypersetup{%
  bookmarksnumbered=true,bookmarksopen=false,%
  plainpages=false,
  linktocpage=true,%
  colorlinks=true,breaklinks=true,%
  linkcolor=\mylinkcolor,citecolor=\mycitecolor,urlcolor=\myurlcolor,%
  pdfpagelayout=OneColumn,%
  pageanchor=true,%
}

\title{Reduced transfer operators \\ for singular difference equations}

\author{Hermann Schulz-Baldes
\\
\\
{\small  Friedrich-Alexander-Universit\"at Erlangen-N\"urnberg}
\\
{\small Department Mathematik, Cauerstr. 11, D-91058 Erlangen, Germany}
\\
{\small Email: schuba@mi.uni-erlangen.de}
}

\date{ }

\newtheorem{theorem}{Theorem}
\newtheorem{proposition}[theorem]{Proposition}

\newcommand{\BM}{{\mathbb B}}
\newcommand{\CM}{{\mathbb C}}
\newcommand{\NM}{{\mathbb N}}
\newcommand{\RM}{{\mathbb R}}

\newcommand{\GM}{{\mathbb G}}
\newcommand{\UM}{{\mathbb U}}

\newcommand{\Aa}{{\cal A}}

\newcommand{\Bb}{{\cal B}}

\newcommand{\Oo}{{\cal O}}
\newcommand{\Tt}{{\cal T}}

\newcommand{\Mm}{{\cal M}}
\newcommand{\Cc}{{\cal C}}
\newcommand{\Jj}{{\cal J}}
\newcommand{\Ii}{{\cal I}}

\newcommand{\Hh}{{\cal H}}

\newcommand{\one}{{\bf 1}}

\newcommand{\spec}{\mbox{\rm spec}}

\newcommand{\Ker}{{\rm Ker}} 
\newcommand{\Ran}{{\rm Ran}}

\newcommand{\diag}{{\rm diag}} 
 
\newcommand{\spa}{{\rm span}}

\newcommand{\ess}{{\mbox{\rm\tiny ess}}}

\newcommand{\tot}{{\mbox{\rm\tiny tot}}}
\newcommand{\Res}{R}

\newcommand{\refe}{{\mbox{\rm\tiny rbc}}}
 
\newcommand{\Length}{N} 
\newcommand{\Size}{L} 
\newcommand{\Transfer}{{\mathcal T}}

\begin{document}

\maketitle

\begin{abstract}
For tridiagonal block Jacobi operators, the standard transfer operator techniques only work if the off-diagonal entries are invertible. Under suitable assumptions on the range and kernel of these off-diagonal operators which assure a homogeneous minimal coupling between the blocks, it is shown how to construct reduced transfer operators that have the usual Krein space unitarity property and also a crucial monotonicity in the energy variable. This allows to extend the results of oscillation theory to such systems.

\vspace{.1cm}

\noindent {\bf Keywords:} singular difference equations,  transfer operators, oscillation theory






\end{abstract}



\vspace{1cm}

\section{Introduction}
\label{sec-Intro}

For many decades and both by the {physics and mathematics community, transfer matrices have been} used very efficiently for the study of one-dimensional discrete Schr\"odinger operators, also called Jacobi matrices \cite{Atk,Tes,DEH}. Many of the strictly one-dimensional techniques have been extended to block Jacobi matrices \cite{Atk,SB1,SB2,DEH} and even operator-valued tridiagonal Hamiltonians \cite{ASV,GSV}. The standard transfer matrix can then only be defined if the off-diagonal block entries are invertible (this is briefly reviewed in Section~\ref{sec-Jacobi} below). However, in numerous models this invertibility simply does not hold. As to a concrete example, one may think of a one-dimensional discrete Schr\"odinger operator with a periodic potential of period $L$; then it is natural to consider blocks of size $L$ and an associated $1$-periodic block Jacobi operator; however, because the discrete Laplacian only has next nearest hoping terms, the off-diagonal entries $L\times L$ entries of block Jacobi operator have merely rank $1$ and are therefore not invertible. In this latter case one can work with the monodromy matrix, but many other examples of block Jacobi operators with non-invertible off-diagonal entries arise in the study of two-dimensional topological insulators and one is forced to address this issue \cite{DC,MKY,MK}. The unit cells in these systems are sufficiently large so that not every point in the cells is directly connected to the outside. Further examples of this type are chains of coupled quantum dots. {Let us also mention that tridiagonal block matrices (albeit not selfadjoint) with rank $1$ off-diagonal entries naturally appear in relation to moment problems \cite{KL,DD}, and these works then also construct associated $2\times 2$ matrices.}

\vspace{.2cm}

This note shows how to set up a transfer matrix formalism in such systems, provided three natural hypothesis spelled out in Section~\ref{sec-Construction} hold. The newly constructed reduced transfer operators then have many of the well-known structural properties that one is accustomed to: they are analytic in the complex energy variable (Proposition~\ref{prop-AnalyticExtend});  they allow to construct the solutions of the Schr\"odinger equation given by the three-term recurrence relation (Proposition~\ref{prop-RedTransferSol}); they conserve a Krein space structure which for real operators reduces to the symplectic nature of the transfer operators (Proposition~\ref{prop-IUnitarity}) and finally they have a positivity property w.r.t. the energy variable (Proposition~\ref{prop-Posititvity}). All these facts are of algebraic nature. They allow to extend oscillation theory in energy, initially discovered by Bott for matrix-valued Sturm-Liouville operators \cite{Bot}, to the systems satisfying the three hypothesis (Theorem~\ref{theo-Jacobi}).

\vspace{.2cm}

The main source of inspiration for this investigation was the work of Dwivedi and Chua \cite{DC} which addressed exactly the same issue and was later on extended in \cite{KD,MKY,MK}. Due to a crucial difference in the definition, the generalized transfer operators in \cite{DC} are in general not $\Ii$-unitary (in the sense described below), an important short-coming already noted by the authors. Moreover, they did not show the (non-trivial) analyticity of the reduced transfer operators, nor analyze the monotonicity property in energy and oscillation theory.

\vspace{.2cm}

To keep this note short, only the example of the reduced transfer matrix for a periodic discrete Schr\"odinger operator is worked out  in detail in Section~\ref{sec-Periodic} where it is also shown that the reduced transfer matrix in this example is actually equal to the monodromy matrix. The interested reader can readily write out the reduced transfer operators for other block Jacobi operators, in particular, stemming from models describing topological insulators and semimetals  as in \cite{DC,MKY,MK}. 

\section{Jacobi operators and their transfer operators}
\label{sec-Jacobi}

A Jacobi operator of finite length ${{\Length}}\geq 3$ is of the tridiagonal form
\begin{equation}
\label{eq-matrix}
H_{\Length}
\;=\;
\left(
\begin{array}{ccccccc}
V_1       & T_2  &        &        &         &        \\
T_2^*      & V_2    &  T_3  &        &         &        \\
            & T_3^* & V_3    & \ddots &         &        \\
            &        & \ddots & \ddots & \ddots  &        \\
            &        &        & \ddots & V_{{\Length}-1} & T_{\Length}   \\
     &        &        &        & T_{\Length}^*  & V_{\Length}
\end{array}
\right)
\;,
\end{equation}
where $(V_n)_{n=1,\ldots,{\Length}}$ are bounded selfadjoint operators on Hilbert spaces $(\Hh_n)_{n=1,\ldots,{\Length}}$, $(T_n)_{n=2,\ldots,{\Length}}$ are bounded operators $T_n:\Hh_n\to\Hh_{n-1}$ and all missing block entries of the matrix are $0$-operators. Hence $H_{\Length}$ is a bounded selfadjoint operator on $\Hh_\tot=\oplus_{n=1}^{\Length}\Hh_n$. Periodized versions Jacobi operators have extra entries $T_1^*$ and $T_1$ in the upper right and lower left corner respectively.  A standard situation considered in most works is that all $\Hh_n$ are all isomorphic to the same finite dimensional Hilbert space $\CM^{\Size}$. In this case $H_{\Length}$ is also called (finite) matrix-valued Jacobi matrix or block Jacobi matrix. If, moreover, $L=1$, one simply speaks of a Jacobi matrix. Let us stress that also infinite dimensional fiber spaces $\Hh_n$ appear naturally in the study of higher dimensional systems \cite{ASV,DC,GSV,MKY}. Furthermore, it is also possible to consider unbounded matrix entries under suitable domain assumptions ({\it e.g.} the $V_n$'s are selfadjoint, all with the same domain that is left invariant under the $T_n$'s).

\vspace{.2cm}

The crucial problem also studied in this note is to compute the spectrum of $H_{\Length}$, namely to find those $E\in\RM$ for which there exists a non-vanishing state $\psi^E=(\psi_n^E)_{n=1,\ldots,{\Length}}\in\Hh_\tot$ such that the Schr\"odinger equation holds:
\begin{equation}
\label{eq-JacobiSchro}
H_{\Length}\psi^E
\;=\;
E\,\psi^E
\;.
\end{equation}
The tridiagonal form of $H_{\Length}$ allows to rewrite this as the set of equations
\begin{equation}
\label{eq-ThreeTermFinite}
T_{n+1}\psi^E_{n+1}\,+\,V_n\psi^E_n\,+\,T_n^*\psi^E_{n-1}
\;=\;
E\psi^E_n
\;,
\qquad
{n=2,\ldots,N-1\,,}
\end{equation}
together with the (Dirichlet) boundary conditions
\begin{equation}
\label{eq-BCondLeft}
T_2\psi^E_2\,+\,V_1\psi^E_1
\;=\;
E\psi^E_1
\;,
\end{equation}
and
\begin{equation}
\label{eq-BCondRight}
V_{\Length}\psi^E_{\Length}\,+\,T_{{\Length}}^*\psi^E_{{\Length}-1}
\;=\;
E\psi^E_{\Length}
\;.
\end{equation}
The equation \eqref{eq-ThreeTermFinite} is also called the three-term recurrence relation because $\psi^E_{n+1}$ can be computed from $\psi^E_n$ and $\psi^E_{n-1}$. In particular, if two neighboring values are known, then all others can be computed. One typically starts out at the left end with an initial condition $\psi^E_1\in\Hh_1$, then computes $\psi^E_2$ from \eqref{eq-BCondLeft} and consequently the other fibers from \eqref{eq-ThreeTermFinite}. Clearly this requires the invertibility of the $T_n$. 

\vspace{.2cm}

Under the crucial assumption that the $T_n$ are invertible {(which implies that all $\Hh_n$, $n=1,\ldots,N$, are isomorphic to some given Hilbert space $\Hh$)}, this standard procedure allows to construct $\psi^E$ via the transfer operators ${\Transfer}_n^E$ from $\Hh_{n-1}\oplus\Hh_{n-1}$ to $\Hh_n\oplus\Hh_n$ defined by
\begin{equation}
\label{eq-transfer}
{\Transfer}_n^E
\;=\;
\begin{pmatrix}
(E\,{\bf 1}\,-\,V_n)\,T_n^{-1} & - T_n^* \\
T_n^{-1} & 0
\end{pmatrix}
\;,
\qquad
n=1,\ldots,{\Length}
\;,
\end{equation}
{where $\one:\Hh_{n-1}\to\Hh_n$ is the identity and} $T_1=\one$ (if the periodic case is considered, then one rather uses the extra entry in the lower left corner of \eqref{eq-matrix}, see \cite{SB2}).  More precisely, setting 
$$
\Psi^E_n
\;=\;
\begin{pmatrix}
T_{n+1}\psi^E_{n+1} \\ \psi^E_n
\end{pmatrix}
\;\in\;
\Hh_n\oplus\Hh_n
\;,
$$
one then has
\begin{equation}
\label{eq-TMatrixEqFinite}
\Psi^E_n
\;=\;
{\Transfer}^E_n\,
\Psi^E_{n-1}
\;,
\end{equation}
together with the initial condition $\Psi^E_0\in\Hh_1\oplus 0$, chosen such that the first boundary condition in \eqref{eq-BCondLeft} is automatically satisfied. The upper equation of \eqref{eq-TMatrixEqFinite} is then the three-term recurrence relation \eqref{eq-ThreeTermFinite}, while the lower equation of \eqref{eq-TMatrixEqFinite} is tautological. Note that the equations \eqref{eq-TMatrixEqFinite} allow to construct $\Psi^E_n$ iteratively and therefore, in particular, $\Psi^E_N$. However, this vector does not necessarily satisfy $\Psi^E_N\in 0\oplus\Hh_N$ which is equivalent to the second boundary condition in \eqref{eq-BCondRight}. To determine for which energies $E$ this right boundary condition is satisfied is the object of intersection theory of Lagrangian planes, that is the theory of the Bott-Maslov index. Also let us note that the transfer matrix $\Tt^E_n$ can be defined by the same formula \eqref{eq-transfer} for complex $E\in\CM$. Then the solutions can be constructed from the initial condition $\Psi^E_{0}=\binom{\psi^E_1}{0}$ via \eqref{eq-TMatrixEqFinite}, but of course the right boundary condition \eqref{eq-BCondRight} cannot be satisfied if $E\in\CM\setminus\RM$.

\vspace{.2cm}

In connection with intersection theory and the eigenvalue computation via oscillation theory (see \cite{Bot,SB1,DSW} or Section~\ref{sec-EV} below), it is of crucial importance that  {one has 
$$
({\Transfer}_n^{{\overline{E}}})^*\,\Ii_{n}\,{\Transfer}_n^E
\;=\;
\Ii_{n-1}
\;,
$$
where $\overline{E}$ denotes the complex conjugate of $E$ and $\Ii_n$ is the skew-adjoint operator
$$
\Ii_n
\;=\;
\begin{pmatrix}
0 & -\one
\\
\one & 0
\end{pmatrix}
$$
on $\Hh_n\oplus\Hh_n$. Supposing that $\Hh_n=\Hh$ for all $n=1,\ldots,\Length$, one can then drop the index $n$ on $\Ii_n$. For $E\in\RM$, the above identity means by definition that $\Tt^E_n$ is $\Ii$-unitarity on the Krein space $(\Hh\oplus\Hh,\Ii)$.} It is then well-known that the $\Ii$-unitary operators form a group denoted by 
$$
\UM(\Hh\oplus\Hh,\Ii)
\;=\;
\big\{\Tt\in\BM(\Hh\oplus\Hh)\,:\,\Tt^*\Ii\Tt=\Ii\big\}
\;.
$$

\section{Construction of the reduced transfer operator}
\label{sec-Construction}

In many situations, the off-diagonal entries $T_n$ are not invertible operators. It is the object of this note to show how one can nevertheless construct reduced transfer operators and use them for the study of the solutions of the Schr\"odinger equation. This section introduces the objects involved in the construction and states the necessary hypothesis. 

\vspace{.2cm}

Associated to the $T_n$ are always two subspace of $\Hh_n$, namely $\Ran(T_n^*)$ and $\Ran(T_{n+1})$.

\vspace{.2cm}

\noindent {\bf Hypothesis 1:} {\sl $\Ran(T_n^*)$ is orthogonal to $\Ran(T_{n+1})$ in $\Hh_n$ for all $n=2,\ldots,\Length-1$.}

\vspace{.1cm}

\noindent {\bf Hypothesis 2:} {\sl For all $n=2,\ldots,\Length$, $\Ran(T_n)$ and $\Ran(T_n^*)$ are closed and all isomorphic to a reference Hilbert space $\widehat{\Hh}$.}

\vspace{.2cm}

As $\Ran(T_n^*)=\Ker(T_n)^\perp$, Hypothesis 1 is equivalent to $\Ran(T_n^*)\subset\Ker(T^*_{n+1})$. One can introduce for $n=1,\ldots,\Length$
$$
\Hh_n^-\;=\;\Ran(T_n^*)\;,
\qquad
\Hh_n^+\;=\;\Ran(T_{n+1})\;,
\qquad
\Hh_n^0
\;=\;
\big(\Hh_n^-\oplus\Hh_n^+\big)^\perp
\;,
$$
where for $n=1$ one sets $\Hh_1^-=\{0\}$ and for $n=\Length$ rather $\Hh_{\Length}^+=\{0\}$. Then
\begin{equation}
\label{eq-Grading}
\Hh_n
\;=\;
\Hh_n^-\oplus\Hh_n^0\oplus\Hh_n^+
\;,
\qquad
n=1,\ldots,\Length
\;.
\end{equation}
It is hence natural to introduce the surjective partial isometries $\pi_n^\pm:\Hh_n\to\Hh_n^\pm$ and $\pi_n^0:\Hh_n\to\Hh_n^0$ onto these subspaces. Then $\pi_n^\pm(\pi_n^\pm)^*=\one_{\Hh_n^\pm}$ and  $\pi_n^0(\pi_n^0)^*=\one_{\Hh_n^0}$ are the identity operators, while  $(\pi_n^\pm)^*\pi_n^\pm$ and $(\pi_n^0)^*\pi_n^0$ are the projections {in $\Hh_n$ onto the subspaces} $\Hh_n^\pm$ and $\Hh_n^0$ respectively. One has
\begin{equation}
\label{eq-HnDecomp}
(\pi_n^-)^*\pi_n^-\,+\,(\pi_n^0)^*\pi_n^0\,+\,(\pi_n^+)^*\pi_n^+
\;=\;
\one_{\Hh_n}
\;.
\end{equation}
Note that
\begin{equation}
\label{eq-TnProp}
(\pi_{n-1}^+)^*\pi_{n-1}^+\,T_n\,(\pi_n^-)^*\pi_n^-
\;=\;
T_n
\;,
\qquad
n\,=\,2,\ldots,N
\;.
\end{equation}
One can now introduce the reduced hopping operators by
$$
\widehat{T}_n
\;=\;
\pi_{n-1}^+\,T_n\,(\pi_n^-)^*\,:\,\Hh_n^-\to\Hh_{n-1}^+
\;,
\qquad
n\,=\,2,\ldots,N\;.
$$
By construction, these are invertible operators. If one identifies $\Hh_{n}^-$ and $\Hh_{n-1}^+$ both with $\widehat{\Hh}$, then $\widehat{T}_n$ can be understood as an invertible operator on $\widehat{\Hh}$. Note that one clearly has
$$
T_n\;=\;(\pi^+_{n-1})^*\,\widehat{T}_n\, \pi^-_n
\;,
\qquad
n\,=\,2,\ldots,N\;.
$$

\vspace{.2cm}

Next let us start from the three-term recurrence relation \eqref{eq-ThreeTermFinite}. Setting, for $n=1,\ldots,N$,
\begin{equation}
\label{eq-PsiRedDef}
\psi^{E,\pm}_n\;=\;\pi_n^\pm\psi^E_n\,\in\,\Hh_n^\pm\,\cong\,\widehat{\Hh}
\;,
\qquad
\psi^{E,0}_n\;=\;\pi_n^0\psi^E_n\,\in\,\Hh_n^0
\;,
\end{equation}
it becomes, for $n=2,\ldots,N-1$,
$$
(\pi^+_{n})^*\widehat{T}_{n+1}\psi^{E,-}_{n+1}
\;=\;
(E\one-V_n)\psi^{E}_n
\,-\,
(\pi^-_{n})^*(\widehat{T}_{n})^*\psi^{E,+}_n
\;.
$$
The next step is  to invert the operator $E\one-V_n$ on $\Hh_n$. This is possible if $E$ is not in the spectrum $\spec(V_n)$ of $V_n$. As this needs to be done for all $n=1,\ldots, N$ (also the boundary cases $n=1$ and $n=N$), it is useful to introduce a notation for the joint resolvent set:
$$
\Res\;=\;\bigcap_{n=1,\ldots,N} \Res_n\;,
\qquad
\Res_n\;=\;\CM\setminus\spec(V_n)
\;.
$$
For $E\in\Res$, one can apply the inverse $(E\one-V_n)^{-1}$ from the left to obtain
$$
(E\one-V_n)^{-1}(\pi^+_{n})^*\widehat{T}_{n+1}\psi^{E,-}_{n+1}
\;=\;
\psi^{E}_n
\,-\,
(E\one-V_n)^{-1}(\pi^-_{n})^*(\widehat{T}_{n})^*\psi^{E,+}_n
\;.
$$
Applying from the left $\pi^-_n $,  $\pi^0_n $ and $\pi^+_n $ results in 
\begin{align*}
\pi^-_n (E\one-V_n)^{-1}(\pi^+_{n})^*\widehat{T}_{n+1}\psi^{E,-}_{n+1}
&
\;=\;
\psi^{E,-}_n
\,-\,
\pi^-_n (E\one-V_n)^{-1}(\pi^-_{n})^*(\widehat{T}_{n})^*\psi^{E,+}_n
\;,
\quad n=2,\ldots,N\,,
\\
\pi^0_n (E\one-V_n)^{-1}(\pi^+_{n})^*\widehat{T}_{n+1}\psi^{E,-}_{n+1}
&
\;=\;
\psi^{E,0}_n
\,-\,
\pi^0_n (E\one-V_n)^{-1}(\pi^-_{n})^*(\widehat{T}_{n})^*\psi^{E,+}_n
\;,
\quad n=1,\ldots,N\,,
\\
\pi^+_n (E\one-V_n)^{-1}(\pi^+_{n})^*\widehat{T}_{n+1}\psi^{E,-}_{n+1}
&
\;=\;
\psi^{E,+}_n
\,-\,
\pi^+_n (E\one-V_n)^{-1}(\pi^-_{n})^*(\widehat{T}_{n})^*\psi^{E,+}_n
,
\quad n=1,\ldots,N-1.
\end{align*}
Note that for $n=1$, the first equation is absent, and for $n=N$ the last one, simply because $\pi^-_1=0$ and $\pi^+_N=0$. Due to the above, it is hence useful to introduce notations for this inverse in the grading of $\Hh_n=\Hh_n^-\oplus\Hh_n^0\oplus\Hh_n^+$:
$$
(E\one-V_n)^{-1}
\;=\;
\begin{pmatrix}
G^{E,-,-}_n & G^{E,-,0}_n & G^{E,-,+}_n
\\
G^{E,0,-}_n & G^{E,0,0}_n & G^{E,0,+}_n
\\
G^{E,+,-}_n & G^{E,+,0}_n & G^{E,+,+}_n
\end{pmatrix}
\;,
\qquad
n=2,\ldots,N-1\,,
$$
namely $G^{E,\pm,-}_n=\pi^\pm_n(E\one-V_n)^{-1}(\pi^-_n)^*$, etc.. For $n=1$ and $n=N$, $(E\one-V_n)^{-1}$ only is a $2\times 2$ matrix. With these notations, one gets the equations
\begin{align}
\label{eq-InvIntermed-}
G^{E,-,+}_n \widehat{T}_{n+1}\psi^{E,-}_{n+1}
& \;=\;
\psi^{E,-}_n
\,-\,
G^{E,-,-}_n (\widehat{T}_{n})^*\psi^{E,+}_n
\;,
\qquad
n=2,\ldots,N-1\,,
\\
\label{eq-InvIntermed0}
G^{E,0,+}_n \widehat{T}_{n+1}\psi^{E,-}_{n+1}
& \;=\;
\psi^{E,0}_n
\,-\,
G^{E,0,-}_n (\widehat{T}_{n})^*\psi^{E,+}_n
\;,\qquad\;
n=2,\ldots,N-1\,,
\\
\label{eq-InvIntermed+}
G^{E,+,+}_n \widehat{T}_{n+1}\psi^{E,-}_{n+1}
& 
\;=\;
\psi^{E,+}_n
\,-\,
G^{E,+,-}_n (\widehat{T}_{n})^*\psi^{E,+}_n
\;,\qquad
n=1,\ldots,N-1\,
.
\end{align}
Now equations \eqref{eq-InvIntermed-} and \eqref{eq-InvIntermed+} are linear relations between $\psi^{E,-}_n$ and $\psi^{E,+}_n$, $n=1,\ldots,N$. Similar as in \eqref{eq-TMatrixEqFinite}, this can be rewritten as
\begin{align*}
&
\begin{pmatrix}
\widehat{T}_{n+1}\psi^{E,-}_{n+1} \\ \psi^{E,+}_n
\end{pmatrix}
\\
&\;\; =\,
\begin{pmatrix}
(G^{E,-,+}_n)^{-1}(\widehat{T}_{n})^{-1} & -(G^{E,-,+}_n)^{-1}G^{E,-,-}_n(\widehat{T}_{n})^*
\\
G^{E,+,+}_n(G^{E,-,+}_n)^{-1}(\widehat{T}_{n})^{-1} & (G^{E,+,-}_n-G^{E,+,+}_n ( G^{E,-,+}_n)^{-1}G^{E,-,-}_n  )(\widehat{T}_{n})^*
\end{pmatrix}
\begin{pmatrix}
\widehat{T}_{n}\psi^{E,-}_{n} \\ \psi^{E,+}_{n-1}
\end{pmatrix}
,
\end{align*}
for $n=2,\ldots,N-1$. Indeed, the first line of this equation is \eqref{eq-InvIntermed-}, and the second line is obtained by replacing \eqref{eq-InvIntermed-} into \eqref{eq-InvIntermed+}. Therefore, one is led to define the reduced transfer operator for $n=2,\ldots,N-1$ as
\begin{equation}
\label{eq-DefRedTransfer}
\widehat{\Tt}^E_n
\; =\;
\begin{pmatrix}
(G^{E,-,+}_n)^{-1}(\widehat{T}_{n})^{-1} & -(G^{E,-,+}_n)^{-1}G^{E,-,-}_n(\widehat{T}_{n})^*
\\
G^{E,+,+}_n(G^{E,-,+}_n)^{-1}(\widehat{T}_{n})^{-1} & (G^{E,+,-}_n-G^{E,+,+}_n ( G^{E,-,+}_n)^{-1}G^{E,-,-}_n  )(\widehat{T}_{n})^*
\end{pmatrix}
\;.
\end{equation}
For the boundary terms, it is convenient to set 
\begin{equation}
\label{eq-DefRedTransferStart}
\widehat{\Tt}^E_1
\;=\;
\begin{pmatrix}
(G^{E,+,+}_1)^{-1}
&
- \one
\\
\one & 0
\end{pmatrix}
\;,
\qquad
\widehat{\Tt}^E_N
\;=\;
\begin{pmatrix}
(G^{E,-,-}_N)^{-1}(\widehat{T}_N)^{-1}
&
- (\widehat{T}_N)^*
\\
(\widehat{T}_N)^{-1} & 0
\end{pmatrix}
\;.
\end{equation}

All the reduced transfer operators $\widehat{\Tt}^E_n$ are by their very definition analytic on the joint resolvent set $\Res$. Furthermore, recall that the discrete spectrum of an operator consists of all isolated eigenvalues of finite multiplicity and the remainder of its spectrum is called the essential spectrum. Here it is useful to introduce the joint essential resolvent set as 
$$
\Res_\ess\;=\;\bigcap_{n=1,\ldots,N} \Res_{n,\ess}\;,
\qquad
\Res_{n,\ess}\;=\;\CM\setminus\spec_\ess(V_n)
\;.
$$
Again by construction, the reduced transfer operators are  meromorphic on $\Res_\ess$, namely they may have poles of finite multiplicity on the discrete spectrum of $V_n$. However, under the following hypothesis, one can verify that these singularities are removable.

\vspace{.2cm}

\noindent {\bf Hypothesis 3:} {\sl  For $n=1,\ldots,N$, if $E$ lies in the discrete spectrum of $V_n$ and $P^{E}_n$ is the orthogonal projection on $\Ker(E\one-V_n)$, the operators $\pi^\pm_nP^{E}_n(\pi_n^\pm)^*:\widehat{\Hh}\to\widehat{\Hh}$ are invertible.}

\vspace{.2cm}

Note that for $n=1$ and $n=N$, this only supposes the invertibility of $\pi^+_1P^{E}_n(\pi_1^+)^*$ and $\pi^-_NP^{E}_N(\pi_N^-)^*$. Let us furthermore point out that due to $\pi^\pm_nP^{E}_n(\pi_n^\pm)^*= (\pi^\pm_nP^{E}_n)(\pi_n^\pm P^{E}_n)^*$ the hypothesis also implies the invertibility $\pi^+_nP^{E}_n(\pi_n^-)^*$.

\begin{proposition}
\label{prop-AnalyticExtend}
Suppose that {\rm Hypothesis 1-3} hold. For $n=1,\ldots,N$ and $E$ in the discrete spectrum of $V_n$, the limit
$$
\widehat{\Tt}^{E}_n
\;=\;
\lim_{e\to E}\,\widehat{\Tt}^{e}_n
$$
exists. The map $E\in\Res_\ess\mapsto \widehat{\Tt}^{E}_n$ is analytic. 
\end{proposition}

\noindent {\bf Proof.} Let us first focus on $n=2,\ldots,N-1$  and rewrite the reduced transfer operator in a factorized form:
\begin{equation}
\label{eq-Factorized}
\widehat{\Tt}^E_n
\;=\;
\begin{pmatrix}
(G^{E,-,+}_n)^{-1} & -(G^{E,-,+}_n)^{-1}G^{E,-,-}_n
\\
G^{E,+,+}_n(G^{E,-,+}_n)^{-1}& G^{E,+,-}_n-G^{E,+,+}_n ( G^{E,-,+}_n)^{-1}G^{E,-,-}_n 
\end{pmatrix}
\begin{pmatrix}
(\widehat{T}^-_{n})^{-1} & 0
\\
0 & (\widehat{T}^-_{n})^*
\end{pmatrix}
\;.
\end{equation}
Hence only the entries in the left factor have to be analyzed. By a shift in energy, one can suppose that $E=0\in\spec(V_n)$. Then $V_n=(\one-P^0_n)V_n(\one-P^0_n)$ so that
$$
E\one-V_n
\;=\;
(\one-P^0_n)(E\one-V_n)(\one-P^0_n)\;+\;E\,P^0_n
\;.
$$
The first summand on the r.h.s. is invertible as an operator on the invariant subspace $\Ran(\one-P^0_n)$. Therefore
\begin{equation}
\label{eq-EExpand}
(E\one-V_n)^{-1}
\;=\;
\big((\one-P^0_n)(E\one-V_n)(\one-P^0_n)\big)^{-1}(\one-P^0_n)\;+\;\frac{1}{E}\,P^0_n
\;,
\end{equation}
so that
$$
G^{E,-,+}_n
\;=\;
\pi^-_n\big((\one-P^0_n)(E\one-V_n)(\one-P^0_n)\big)^{-1}(\one-P^0_n)(\pi^+_n)^*\;+\;\frac{1}{E}\,\pi^-_n P^0_n(\pi^+_n)^*
\;.
$$
The first summand is uniformly bounded in $E$ on a neighborhood of $0$ by construction, while the second is invertible by assumption. Therefore one concludes
$$
\lim_{E\to 0}\;
(G^{E,-,+}_n)^{-1}
\;=\;
0
\;.
$$
As to the lower left entry, let us use \eqref{eq-EExpand} multiplied by $E$ twice to obtain
\begin{align*}
G^{E,+,+}_n(G^{E,-,+}_n)^{-1}
\;=\;
\big(\pi^+_n P^0_n(\pi^+_n)^*+\Oo(E)\big)^{-1}
\big(\pi^-_n P^0_n(\pi^+_n)^*+\Oo(E)\big)^{-1}
\;.
\end{align*}
Again the limit exists. The upper right is done is obtained in the same manner. For the lower right corner, note that up to terms of order $\Oo(1)$
$$
G^{E,+,-}_n\,-\,G^{E,+,+}_n ( G^{E,-,+}_n)^{-1}G^{E,-,-}_n 
\,=\,
\frac{1}{E}
\Big[
\pi^+_n P^0_n(\pi^-_n)^*
\,-\,
\pi^+_n P^0_n(\pi^+_n)^*(\pi^+_n P^0_n(\pi^-_n)^*)^{-1}\pi^-_n P^0_n(\pi^-_n)^*
\Big]
\;.
$$
As $\pi^+_n P^0_n(\pi^-_n)^*=(\pi^+_n P^0_n)(\pi^-_nP^0_n)^*$ is invertible, one realizes that the coefficient of $E^{-1}$ actually vanishes. Therefore again the limit $E\to 0$ exists. The cases $n=1$ and $n=N$ are dealt with in a similar manner by expanding $(G^{E,+,+}_1)^{-1}$ and $(G^{E,-,-}_N)^{-1}$. The analyticity now follows from Riemann's theorem on removable singularities. 
\hfill $\Box$

\vspace{.2cm}

Let us note that the proof also provides all entries of $\widehat{\Tt}^E_n$ except for the lower right one that is a bit more cumbersome to compute. Furthermore, if $V_n$ is a finite-dimensional matrix, then $\Res_\ess=\CM$ and Proposition~\ref{prop-AnalyticExtend} implies that $E\in\CM\mapsto\widehat{\Tt}^E_n$ is an entire function.

\section{Reduced transfer matrices for periodic operators}
\label{sec-Periodic}

This section is a worked out example for the construction of a reduced transfer matrix. Let us consider standard one-dimensional (scalar) periodic Schr\"odinger operator with a periodic potential of period $L$. Hence $H$ is of the form \eqref{eq-matrix} with scalar coefficients $T_n=t_n$ and $V_n=v_n$ satisfying for all $n$
$$
t_n\;=\;1\;,\qquad v_{n+L}\;=\;v_n\;.
$$
For simplicity, let us then assume that $N=KL$ for some $K\in\NM$. It is then natural \cite{Tes} to use the $2\times 2$  monodromy matrix $\Mm^E$ over one period of the potential:
$$
\Mm^E
\;=\;
\begin{pmatrix}
E-v_L & -1 \\ 1 & 0
\end{pmatrix}
\cdots
\begin{pmatrix}
E-v_2 & -1 \\ 1 & 0
\end{pmatrix}
\begin{pmatrix}
E-v_1 & -1 \\ 1 & 0
\end{pmatrix}
\;.
$$
However, another way to look at $H$ is to view it as a $1$-periodic block matrix with $L\times L$ entries, namely to write $H$ as in \eqref{eq-matrix} with $N=K$ and $L\times L$ matrices $V_n$ and $T_n$ that are all equal to
$$
V
\;=\;
\begin{pmatrix}
v_1       & 1  &        &        &                \\
1     & v_2    &  1 &        &                \\
                    & \ddots & \ddots & \ddots  &        \\
                    &        & 1 & v_{L-1} & 1   \\
        &        &        & 1  & v_{L}
\end{pmatrix}
\;,
\qquad
T
\;=\;
\begin{pmatrix}
       &   &        &        &                \\
   &     &   &        &                \\
                    &  &  &   &        \\
                    &        &  &  &    \\
1        &        &        &   & 
\end{pmatrix}
\;,
$$
respectively. Here again all empty entries contain $0$'s. Hence $T$ is a rank $1$ matrix, while $V$ is an $L\times L$ Jacobi matrix with Dirichlet boundary conditions. Once the Hypothesis 1-3 have been checked, one can hence associate reduced transfer matrices $\widehat{\Tt}^E_n$ which for all sites $n=2,\ldots, K-1$ are identical to one and the same reduced transfer matrix denoted by $\widehat{\Tt}^E$.

\vspace{.2cm}

To check the hypothesis, let $e_1,\ldots,e_L$ denote the standard basis of $\CM^L$ (in which $V$ and $T$ are represented above). Then $\Ran(T)=\spa\{e_L\}$ and $\Ran(T^*)=\spa\{e_1\}$. In particular, Hypothesis 1 and 2 hold. Then $\widehat{T}=\widehat{T}_n=1$ are scalar entries. Furthermore, let us note that $V$ is a $L\times L$ Jacobi matrix with Dirichlet boundary conditions on both ends. Its spectrum is known to be simple with eigenvectors that are non-vanishing at the first and last site of $\CM^{L}$, which directly implies Hypothesis 3. In conclusion the constructions of Section~\ref{sec-Construction} produce a $2\times 2$ reduced transfer matrix $\widehat{\Tt}^E$. The results of Section~\ref{sec-Properties} show that it satisfies the same equations as the monodromy matrix. Hence one can conclude $\widehat{\Tt}^E=\Mm^E$. This  shows explicitly that $\widehat{\Tt}^E$ is analytic (what is also assured by Proposition~\ref{prop-AnalyticExtend}).

\section{Structural properties of reduced transfer operators}
\label{sec-Properties}

As a motivation for the construction of the reduced transfer operators $\widehat{\Tt}^E_n$, it was shown in Section~\ref{sec-Construction} that they lead to a reproduction property of the solution to the Schr\"odinger equation. On the other hand, the boundary cases $\widehat{\Tt}^E_1$ and $\widehat{\Tt}^E_N$ were given in an ad hoc manner. Now a more careful analysis of the boundary conditions justifies their definition.

\begin{proposition}
\label{prop-RedTransferSol}
Let $E\in\Res_\ess$ and let {\rm Hypothesis 1-3} hold. Suppose $\psi^E=(\psi^E_n)_{n=1,\ldots,\Length}\in\Hh_\tot$ satisfies the three-term recurrence relation \eqref{eq-ThreeTermFinite} and the left boundary condition \eqref{eq-BCondLeft}. With $\psi^{E,-}_n$, $\psi^{E,0}_n$ and $\psi^{E,+}_n$ as in \eqref{eq-PsiRedDef}, set
$$
\widehat{\Psi}^E_n
\;=\;
\begin{pmatrix}
\widehat{T}_{n+1}\psi^{E,-}_{n+1} \\ \psi^{E,+}_n
\end{pmatrix}
\;\in\;
{\Hh}^+_n\oplus\Hh_n^+\,\cong\,\widehat{\Hh}\oplus\widehat{\Hh}
\;,
\qquad
n=1,\ldots,\Length-1
\;,
$$
as well as
\begin{equation}
\label{eq-InitialCondHat}
\widehat{\Psi}^E_0
\;=\;
\begin{pmatrix}
\psi^{E,+}_{1} \\ 0
\end{pmatrix}
\;.
\end{equation}
Then one has
\begin{equation}
\label{eq-GenTransStates}
\widehat{\Psi}^E_n
\;=\;
\widehat{\Tt}^E_n
\,
\widehat{\Psi}^E_{n-1}
\;,
\qquad
n=1,\ldots,{\Length-1}\;,
\end{equation}
and the middle pieces are given by
\begin{equation}
\label{eq-MiddlePieces}
\psi^{E,0}_n
\;=\;
G^{E,0,+}_n \widehat{T}_{n+1}\psi^{E,-}_{n+1}
\,+\,
G^{E,0,-}_n (\widehat{T}_{n})^*\psi^{E,+}_n
\;,
\qquad
n=1,\ldots,\Length\;,
\end{equation}
where $\psi^{E,-}_1=0$ and $\psi^{E,+}_N=0$. 

\vspace{.1cm}

Inversely, one can start out with an arbitrary initial condition $\widehat{\Psi}^E_0$ of the form given in \eqref{eq-InitialCondHat}, namely the lower component vanishes. One then constructs $\widehat{\Psi}^E_n$, $n=1,\ldots,\Length$ by \eqref{eq-GenTransStates}. Using only $\widehat{\Psi}^E_1,\ldots,\widehat{\Psi}^E_{N-1}$ one then obtains $\psi^{E,0}_n$ by \eqref{eq-MiddlePieces} and consequently $\psi^E_n=(\psi^{E,-}_n,\psi^{E,0}_n,\psi^{E,+}_n)$. Then $\psi^E=(\psi^E_n)_{n=1,\ldots,\Length}$ satisfies the three-term recurrence relation \eqref{eq-ThreeTermFinite} and the left boundary condition \eqref{eq-BCondLeft}. 

\vspace{.1cm}

Furthermore, $\psi^E=(\psi^E_n)_{n=1,\ldots,\Length}$ also satisfies the right boundary condition \eqref{eq-BCondRight} and hence is an eigenvector with eigenvalue $E$ if and only if
\begin{equation}
\label{eq-EigenvalueCond}
\widehat{\Tt}^E_N
\,
\widehat{\Psi}^E_{N-1}
\;=\;
\begin{pmatrix}
0 \\ \psi^{E,-}_N
\end{pmatrix}
\;.
\end{equation}
\end{proposition}

\noindent {\bf Proof.} Due to the analyticity of all objects involved (see Proposition~\ref{prop-AnalyticExtend}), it is sufficient to prove the statement under the assumption that all inverses exist, as it then also follows for all $E\in\Res_\ess$ by analytic continuation.  After equations \eqref{eq-InvIntermed-} to \eqref{eq-InvIntermed+}, it was already shown that \eqref{eq-GenTransStates} holds for $n=2,\ldots,N-1$. Moreover, \eqref{eq-InvIntermed0}  provides \eqref{eq-MiddlePieces} also for $n=2,\ldots,N-1$. For $n=1$, let us note that \eqref{eq-BCondLeft} becomes
$$
(E\one-V_1)^{-1}(\pi^+_{1})^*\widehat{T}_{2}\psi^{E,-}_{2}
\;=\;
\psi^{E}_1
\;,
$$
so that
$$
G^{E,0,+}_1 \widehat{T}_{2}\psi^{E,-}_{2}\;=\;\psi^{E,0}_1\;,
\qquad
G^{E,+,+}_1 \widehat{T}_{2}\psi^{E,-}_{2}\;=\;\psi^{E,+}_1\;.
$$
The former equation is just \eqref{eq-MiddlePieces} for $n=1$, and the latter equation can be rewritten as
$$
\begin{pmatrix}
\widehat{T}_{2}\psi^{E,-}_{2} \\ \psi^{E,+}_{1}
\end{pmatrix}
\;=\;
\begin{pmatrix}
(G^{E,+,+}_1)^{-1} & -\one \\ \one & 0
\end{pmatrix}
\begin{pmatrix}
\psi^{E,-}_{1} \\ 0
\end{pmatrix}
\;,
$$
which is precisely \eqref{eq-GenTransStates} for $n=1$ due to the definition of $\widehat{\Tt}^E_1$. 

\vspace{.1cm}

Furthermore, \eqref{eq-BCondRight} becomes
$$
(E\one-V_N)^{-1}(\pi^-_{N})^*(\widehat{T}_{N})^*\psi^{E,+}_{N-1}
\;=\;
\psi^{E}_N
\;,
$$
so that
$$
G^{E,-,-}_N (\widehat{T}_N)^*\psi^{E,-}_{N-1}\;=\;\psi^{E,-}_N\;,
\qquad
G^{E,0,-}_N (\widehat{T}_N)^*\psi^{E,-}_{N-1}\;=\;\psi^{E,0}_N\;.
$$
The first equation can be rewritten as
$$
\begin{pmatrix}
0 \\ \psi^{E,+}_{N}
\end{pmatrix}
\;=\;
\begin{pmatrix}
(G^{E,-,-}_N)^{-1}(\widehat{T}_{N})^{-1} & - (\widehat{T}_{N})^*\\ (\widehat{T}_{N})^{-1} & 0
\end{pmatrix}
\begin{pmatrix}
\widehat{T}_{N}\psi^{E,-}_{N} \\ \psi^{E,-}_{N-1} 
\end{pmatrix}
\;,
$$
which is \eqref{eq-EigenvalueCond} due to the definition of $\widehat{\Tt}^E_N$, and the second equation is the case $n=N$ of  \eqref{eq-MiddlePieces}. All computations can be read up-side down, also showing the inverse implication.
\hfill $\Box$

\vspace{.2cm}

The next result shows that the reduced transfer matrix has the usual symmetry property. 

\begin{proposition}
\label{prop-IUnitarity}
Suppose that $E\in \Res_\ess$ and that {\rm Hypothesis 1-3} hold. Then the reduced transfer operators defined by \eqref{eq-DefRedTransfer} and \eqref{eq-DefRedTransferStart} satisfy the generalized $\widehat{\Ii}$-unitary relation
$$
(\widehat{\Tt}^{\overline{E}}_n)^*\,\widehat{\Ii}\,\widehat{\Tt}^{E}_n
\;=\;
\widehat{\Ii}
\;,
\qquad
\widehat{\Ii}
\;=\;
\begin{pmatrix}
0 & -\one
\\
\one & 0
\end{pmatrix}
\;.
$$
In particular, for $E\in\RM\cap\Res_\ess$ all these reduced transfer operators are $\widehat{\Ii}$-unitary.
\end{proposition}

\noindent  {\bf Proof.} Let us introduce the following notations for the two factors on the r.h.s. in \eqref{eq-Factorized}:
$$
\Aa^E_n
\,=\,
\begin{pmatrix}
(G^{E,-,+}_n)^{-1} & -(G^{E,-,+}_n)^{-1}G^{E,-,-}_n
\\
G^{E,+,+}_n(G^{E,-,+}_n)^{-1}& G^{E,+,-}_n-G^{E,+,+}_n ( G^{E,-,+}_n)^{-1}G^{E,-,-}_n 
\end{pmatrix}
\;,
\quad
\Bb_n
\,=\,
\begin{pmatrix}
(\widehat{T}_{n})^{-1} & 0
\\
0 & (\widehat{T}_{n})^*
\end{pmatrix}
,
$$
namely $\widehat{\Tt}^{E}_n=\Aa^E_n\Bb_n$. For \eqref{eq-DefRedTransfer} it is sufficient to show that for operators satisfy the generalized $\widehat{\Ii}$-unitary relation. It is straight-forward to algebraically check $(\Bb_n)^* \widehat{\Ii}\Bb_n=\widehat{\Ii}$. As to $\Aa^E_n$, one has
\begin{equation}
\label{eq-AI}
(\Aa^{\overline{E}}_n)^*\,\widehat{\Ii}
\;=\;
\begin{pmatrix}
(G^{E,+,-}_n)^{-1}G^{E,+,+}_n & -(G^{E,+,-}_n)^{-1}
\\
G^{E,-,+}_n-G^{E,-,-}_n ( G^{E,+,-}_n)^{-1}G^{E,+,+}_n
 & G^{E,-,-}_n(G^{E,+,-}_n)^{-1}
\end{pmatrix}
\;,
\end{equation}
and a short computation then shows $(\Aa^{\overline{E}}_n)^*\widehat{\Ii}\Aa^{{E}}_n=\widehat{\Ii}$. For the operators in \eqref{eq-DefRedTransferStart} this is also merely an algebraic computation. 
\hfill $\Box$

\vspace{.2cm}

Let us note that by a non-linear transformation, one can associated to the reduced transfer operator also a reduced scattering operator which is then a unitary operator on $\widehat{\Hh}\oplus\widehat{\Hh}$ \cite[Remark 6]{BFS}. Instead of spelling this out explicitly, let us rather verify the fundamental positivity property that is of crucial relevance for oscillation theory. 

\begin{proposition}
\label{prop-Posititvity}
Suppose $E\in\RM\cap\Res_\ess$ and that {\rm Hypothesis 1-3} hold. Then the reduced transfer operators defined by \eqref{eq-DefRedTransfer} and \eqref{eq-DefRedTransferStart} satisfy
$$
(\widehat{\Tt}^E_n)^*\,\widehat{\Ii}\,\partial_E
\widehat{\Tt}^E_n
\;\geq\;0
\;,
\qquad
n=1,\ldots,N\;.
$$
\end{proposition}

\noindent {\bf Proof.} By analyticity, it is sufficient to prove the claim for $E\in\Res$, notably when $\widehat{\Tt}^E_n$ is given by \eqref{eq-DefRedTransfer} and \eqref{eq-DefRedTransferStart}. Let us first consider the case $n=2,\ldots,N-1$ and for now drop the indices $n$ and $E$. Hence $\partial$ denotes the derivative w.r.t. $E$. Also recall the notations $\Aa=\Aa^E_n$ and $\Bb=\Bb_n$ from the proof of Proposition~\ref{prop-IUnitarity}, namely $\widehat{\Tt}=\Aa\Bb$. Then $\Bb$ is invertible and it is sufficient to show that
$$
(\Bb^{-1})^*\widehat{\Tt}^*\,\widehat{\Ii}\,
\partial\,\widehat{\Tt}
\Bb^{-1}
\;=\;
\Aa^*\,\widehat{\Ii}\, \partial\,\Aa
$$
is non-negative. Using \eqref{eq-AI} for $\Aa^*\widehat{\Ii}$ and writing out the derivative $\partial\Aa$ explicitly, one finds after a tedious but elementary computation
\begin{align*}
\begin{pmatrix}
G^{-,+}
&
0
\\
0 & \one
\end{pmatrix}^*
&
\Aa^*\,\widehat{\Ii}\, \partial\,\Aa
\begin{pmatrix}
G^{-,+}
&
0
\\
0 & \one
\end{pmatrix}
\\
&
\;=\;
\begin{pmatrix}
-\partial G^{+,+} & \partial G^{+,+} C -\partial G^{+,-}
\\ 
C^*\partial G^{+,+}-\partial G^{-,+} & \partial G^{-,+}C + C^*\partial G^{+,-}  -\partial G^{-,-} -  C^*\partial G^{+,+}C
\end{pmatrix}
\end{align*}
where $C=( G^{-,+})^{-1} G^{-,-}$. Reintroducing temporarily $n$ and $E$, let us define the injective operators
$$
A\,=\,(E\,\one\,-\,V_n)^{-1}(\pi^+_n)^*\,:\,\widehat{\Hh}\to\Hh_n\;,
\qquad
B\,=\,(E\,\one\,-\,V_n)^{-1}(\pi^-_n)^*\,:\,\widehat{\Hh}\to\Hh_n\;.
$$
Then
$$
-\partial G^{+,-}
\;=\;
-\partial\,\pi_n^+(E\,\one\,-\,V_n)^{-1}(\pi^-_n)^*
\;=\;
\pi_n^+(E\,\one\,-\,V_n)^{-2}(\pi^-_n)^*
\;=\;
A^*B\;,
$$
and similarly $-\partial G^{+,+}=A^*A$ as well as $-\partial G^{-,-}=B^*B$. Thus 
\begin{align*}
\begin{pmatrix}
G^{-,+}
&
0
\\
0 & \one
\end{pmatrix}^*
\Aa^*\,\widehat{\Ii}\, \partial\,\Aa
\begin{pmatrix}
G^{-,+}
&
0
\\
0 & \one
\end{pmatrix}
&
\;=\;
\begin{pmatrix}
A^*A & A^*(B-A C)
\\ 
(B-AC)^*A & (B-AC)^*(B-AC)
\end{pmatrix}
\\
&
\;=\;
\begin{pmatrix}
A & 0
\\ 
0 & B-AC
\end{pmatrix}^*
\begin{pmatrix}
\one & \one
\\ 
\one & \one
\end{pmatrix}
\begin{pmatrix}
A & 0
\\ 
0 & B-AC
\end{pmatrix}
\;,
\end{align*}
which is clearly non-negative because the matrix in the middle has spectrum $\{0,2\}$. For $\widehat{\Tt}^E_1$, one has
\begin{align*}
(\widehat{\Tt}^E_1)^*\,\widehat{\Ii}\,
\partial\,\widehat{\Tt}^E_1
&
\;=\;
\begin{pmatrix}
(G^{E,-,-}_1)^{-1}
&
- \one
\\
\one & 0
\end{pmatrix}
^*
\widehat{\Ii}\,
\partial 
\begin{pmatrix}
(G^{E,-,-}_1)^{-1}
&
- \one
\\
\one & 0
\end{pmatrix}
\;=\;
\begin{pmatrix}
\partial(G^{E,-,-}_1)^{-1}
&
0
\\
0 & 0
\end{pmatrix}
\;,
\end{align*}
which due to $\partial(G^{E,-,-}_1)^{-1}=(G^{E,-,-}_1)^{-1}
(-\partial G^{E,-,-}_1) (G^{E,-,-}_1)^{-1}$ is also non-negative. Finally, for $\widehat{\Tt}^E_N$ the same argument applies.
\hfill $\Box$

\section{Oscillation theory in energy}
\label{sec-EV}

In this section, it will be supposed that $\dim(\Hh_n)$ is finite {for all $n=1,\ldots,N$}, but these dimensions need not all be equal. However, as stated in Hypothesis 2, $\dim(\Hh^\pm_n)=\dim(\widehat{\Hh})$ is equal to constant denoted by $L$. Hence we will simply identify $\widehat{\Hh}\cong \CM^{L}$ and, for sake of notational simplicity also denote the Krein space $(\widehat{\Hh}\oplus\widehat{\Hh},\widehat{\Ii})$ simply by $(\CM^{2L},\Ii)$ where ${\Ii}$ is the $2L\times 2L$ matrix given as in Proposition~\ref{prop-IUnitarity}. In the following, the structures on this Krein space will be heavily used and therefore they are briefly reviewed (see {\it e.g.} \cite{SB1,DSW}). The set of $\Ii$-unitary matrices $\UM(\CM^{2L},\Ii)=\{\Tt\in\CM^{2L\times 2L}\,:\,\Tt^*\Ii\Tt=\Ii\}$ is a subgroup of the general linear group $\GM(\CM,2L)$ of invertible $2L\times 2L$ matrices. Via Cayley transform it is isomorphic to the generalized Lorentz group of $\Jj$-unitary matrices $\UM(\CM^{2L},\Jj)$ satisfying $\Tt^*\Jj\Tt=\Jj$ where $\Jj=\diag(\one,-\one)$.   
%
%
An $\Ii$-Lagrangian frame  is a $2L\times L$ matrix  $\Phi$ of full rank $L$ satisfying ${\Phi}^*{\Ii}{\Phi}=0$ (the normalization condition ${\Phi}^*{\Phi}=\one$  is not required here). An ${\Ii}$-Lagrangian subspace of the Krein space is the range of an ${\Ii}$-Lagrangian frame. All frames for a given ${\Ii}$-Lagrangian subspace only differ by a right multiplication by an invertible matrix from $\GM(\CM,L)$. The set of ${\Ii}$-Lagrangian subspaces is isomorphic to $\UM(L)$ by the so-called stereographic projection
$$
\Pi({\Phi})
\;=\;
\binom{\one}{\imath\one}^*{\Phi}\,\left[\binom{\one}{-\imath\one}^*{\Phi}\right]^{-1}
\;.
$$
Note that this is indeed a class map, namely $\Pi({\Phi})=\Pi({\Phi}V)$ for any $V\in\GM(\CM,L)$. If $\Phi$ is an ${\Ii}$-Lagrangian frame and $\Tt$ is an $\Ii$-unitary, then $\Tt\Phi$ is also an ${\Ii}$-Lagrangian frame. Under the stereographic projection, this induces an action of $\UM(\CM^{2L},\Ii)$ on $\UM(L)$ by M\"obius transformation with the Cayley transform of $\Tt$. Finally, let $\Phi$ and $\Phi'$ be two $\Ii$-Lagrangian frames, then the intersection of their ranges can be read off the spectral theory of their stereographic projections:
$$
\dim\big(\Ran(\Phi)\cap\Ran(\Phi')\big)
\;=\;
\dim\big(\Ker\big(\Pi(\Phi')^*\Pi(\Phi)-\one\big)\big)
\;.
$$
While elementary, this fact is at the root of intersection theory of $\Ii$-Lagrangian planes and the theory of the Bott-Maslov index.

\vspace{.2cm}

After these preparatory reminders, let us now come back to the eigenvalue problem \eqref{eq-JacobiSchro} for $H_N$ satisfying Hypothesis~1 to 3 of Section~\ref{sec-Construction}. It will be shown how the reduced transfer matrices can be used, based on Proposition~\ref{prop-RedTransferSol}. For this purpose, let us note that the permitted (Dirichlet) initial conditions $\widehat{\Psi}^E_0$ in \eqref{eq-InitialCondHat} make out the range of an $\Ii$-Lagrangian frame 
$$
\Phi^E_0\;=\;\binom{\one}{0}
\;.
$$
The corresponding $L$-dimensional space of solutions can due to \eqref{eq-GenTransStates} now be obtained iteratively by applying the reduced transfer matrices. As these latter are $\Ii$-unitary by Proposition~\ref{prop-IUnitarity}, one hence obtains a sequence of $\Ii$-Lagrangian frames
$$
\Phi^E_n
\;=\;
\widehat{\Tt}^E_n \Phi^E_{n-1}
\;,
\qquad
n=1,\ldots,N
\;.
$$
The final $\Ii$-Lagrangian frame $\Phi^E_N$ does not necessarily satisfy the right boundary condition in {\eqref{eq-EigenvalueCond}} given by $\Phi_\refe=\binom{0}{\one}$. Note that $\Phi_\refe$ is also an $\Ii$-Lagrangian frame. However, according to Proposition~\ref{prop-IUnitarity} the dimension of the intersection of the two subspaces $\Ran(\Phi^E_N)$ and $\Ran(\Phi_\refe)$ is precisely the multiplicity of $E$ as eigenvalue of $H_N$. Combined with the above, one obtains
\begin{align*}
\dim\big(\Ker(H_N-E)\big)
&
\;=\;
\dim\big(\Ran(\Phi^E_N)\cap\Ran(\Phi_\refe)\big)
\;=\;
\dim\big(\Ker\big(\Pi(\Phi_\refe)^*\Pi(\Phi^E_N)-\one\big)\big)
\;.
\end{align*}
Here one can replace $\Pi(\Phi_\refe)=-\one$. Moreover, it is useful to introduce the (reduced) matrix Pr\"ufer phase at site $N$ and energy $E$ as
$$
\widehat{U}^E_N
\;=\;
\Pi(\Phi^E_N)
\,\in\,\UM(L)
\;.
$$
The first part of the following result is then already clear from the above. The second statement implies that the number up to a given energy can be computed as the spectral flow through $-1$ of the path of unitaries $e\in(-\infty,E]\mapsto \widehat{U}^e_{\Length}$. This spectral flow is also equal to the Bott-Maslov index of $e\in(-\infty,E]\mapsto \Phi^e_{\Length}$. For further details on these claims, the reader is referred to \cite{SB1,DSW}.

\begin{theorem} 
\label{theo-Jacobi} 
Let $N\geq 2$. One has
$$
\dim\big(\Ker(H_N-E)\big)
\;=\;
\dim\big(\Ker(\widehat{U}^E_N+\one)\big)
\;.
$$
Moreover,
\begin{equation}
\label{eq-PosPruef}
\frac{1}{\imath}\; (\widehat{U}_{\Length}^E)^*\partial_E \widehat{U}_{\Length}^E
\;>\;0
\;.
\end{equation}
As a function of the energy $E$, the eigenvalues of $\widehat{U}^{E}_{\Length}$ rotate around the unit circle in the positive sense with non-vanishing velocity.
\end{theorem}

\noindent {\bf Proof.}  For the proof of the positivity, let us introduce ${\Size}\times {\Size}$ matrices $a^E$ and $b^E$ by
\begin{equation}
\label{eq-IJlink}
2^{-\frac{1}{2}}\begin{pmatrix}
\one & -\imath \one \\
\one & \imath \one
\end{pmatrix}
{\Phi}^E_{{\Length}}
\;=\;
\binom{a^E}{b^E}
\;.
\end{equation}
They are invertible and by definition $\widehat{U}^E_{\Length}=a^E(b^E)^{-1}=((a^E)^{-1})^*(b^E)^*$. Now
$$
(\widehat{U}^E_{\Length})^*\,\partial_E\,\widehat{U}^E_{\Length}
\;=\;
((b^E)^{-1})^*
\Bigl[\,
(a^E)^*\partial_Ea^E \,-\,
(b^E)^*\partial_Eb^E
\,\Bigr]
(b^E)^{-1}
\;.
$$
Thus it is sufficient to verify positive definiteness of
$$
\frac{1}{\imath}\;
\Bigl[\,
(a^E)^*\partial_Ea^E \,-\,
(b^E)^*\partial_Eb^E
\,\Bigr]
\;=\;
({\Phi}^E_{{\Length}})^*\,\Ii\,\partial_E {\Phi}^E_{{\Length}}
\;,
$$
where \eqref{eq-IJlink} was used. From the product rule it follows that
$$
\partial_E {\Phi}^E_{{\Length}}
\;=\;
\sum_{n=1}^{\Length}
\;
\widehat{\Transfer}^E_{\Length}\cdots \widehat{\Transfer}^E_{n+1}
\,
\left(\partial_E \widehat{\Transfer}^E_n\right)
\;
\widehat{\Transfer}^E_{n-1} \cdots \widehat{\Transfer}^E_1
\,{\Phi}_0^E
\;.
$$
Due to the $\Ii$-unitarity of $\widehat{\Transfer}^E_n$ verified in Proposition~\ref{prop-IUnitarity}, this implies
$$
({\Phi}^E_{{\Length}})^*\,\Ii\,\partial_E {\Phi}^E_{{\Length}}
\;=\;
\sum_{n=1}^{\Length}
\;({\Phi}_0^E)^*\,
\left(
\widehat{\Transfer}^E_{n-1}\cdots \widehat{\Transfer}^E_1
\right)^*
\,
\bigl(\widehat{\Transfer}^E_n\bigr)^*\,\Ii\,
\bigl(\partial_E \widehat{\Transfer}^E_n\bigr)
\;
\left(
\widehat{\Transfer}^E_{n-1}\cdots \widehat{\Transfer}^E_1
\right)
\,{\Phi}_0^E
\;.
$$
Now Proposition~\ref{prop-Posititvity} implies the non-negativity in the claim. The (strict) positivity follows by showing that the sum of two summands is already strictly positive, by an argument as in \cite{SB1}. This directly implies the last claim.
\hfill $\Box$

\vspace{.2cm}

\noindent {\bf Acknowledgements:} The author thanks the two referees for several comments that lead to a considerable improvement of the manuscript. This work was supported by the DFG grant SCHU 1358/6-2. 


\end{document}